\documentclass[12pt]{article}
\usepackage{graphicx}
\usepackage{xspace}
\usepackage{hyperref}
\newcommand\pubnumber{}
\newcommand\pubdate{\today}

\def\rwththreeA{III. Physikalisches Instiut A \\
Physikzentrum, RWTH Aachen University, 52056 Aachen\\
t.pook $<$at$>$ cern.ch}
\def\support{\footnote{This work is supported by the german Federal Ministry of Education and Research}}

\def\Title#1{\begin{center} {\Large #1 } \end{center}}
\def\Author#1{\begin{center}{ \sc #1} \end{center}}
\def\Address#1{\begin{center}{ \it #1} \end{center}}

\newcommand\pubblock{\rightline{\begin{tabular}{l} \pubnumber\\
         \pubdate  \end{tabular}}}
\newenvironment{Abstract}{\begin{quotation}  }{\end{quotation}}
\newenvironment{Presented}{\begin{quotation} \begin{center} 
             PRESENTED AT\end{center}\bigskip 
      \begin{center}\begin{large}}{\end{large}\end{center} \end{quotation}}
\def\Acknowledgements{\bigskip  \bigskip \begin{center} \begin{large}
             \bf ACKNOWLEDGEMENTS \end{large}\end{center}}
\def\beq{\begin{equation}}
\def\eeq#1{\label{#1}\end{equation}}
\def\eeqn{\end{equation}}

\def\beqa{\begin{eqnarray}}
\def\eeqa#1{\label{#1}\end{eqnarray}}
\def\eeqan{\end{eqnarray}}

\let\bar=\overbar

\def\Dslash{\not{\hbox{\kern-4pt $D$}}}
\def\dslash{\not{\hbox{\kern-2pt $\del$}}}

\def\msb{{\bar{\ssstyle M \kern -1pt S}}}

\def\GeV{\,\textrm{GeV}\xspace}
\def\TeV{\,\textrm{TeV}\xspace}
\def\Wprime{\ensuremath{\mathrm{W'}}}
\def\Zprime{\ensuremath{\mathrm{Z^\prime}}}
\def\met{\ensuremath{{\not\mathrel{E}}_T}\xspace}
\def\boldmet{\ensuremath{\not\mathrel{\textbf{E}}}_{\small\textbf{T}}} 
\begin{document}
\begin{titlepage}
\pubblock

\vfill
\Title{Extra Dimension-Inspired Models: \Zprime, \Wprime, Dijet Resonances, Black Hole Searches}
\vfill
\Author{ Tobias Pook on behalf of the ATLAS \& CMS Collaborations\support}
\Address{\rwththreeA}
\vfill
\begin{Abstract}
I give a summary of BSM searches performed by the ATLAS and CMS experiments 
with an focus on heavy gauge bosons, extra dimensions and quantum 
black holes. The presented results use data collected during 2012 when the
LHC operated at an center of mass energy of $\sqrt{s}=8\TeV$.{}
\\

\centering \textit{In memory of Harris Hassan}
\end{Abstract}
\vfill
\begin{Presented}
the Twelfth Conference on the Intersections of Particle and Nuclear Physics (CIPANP15)\\
Vail, USA,  May 19--24, 2015
\end{Presented}
\vfill
\end{titlepage}
\def\thefootnote{\fnsymbol{footnote}}
\setcounter{footnote}{0}

\section{Introduction}

The Large Hadron Collider (LHC) operated at a center of mass energy of 
$\sqrt{s}=8\,\mathrm{TeV}$ during 2012 and the multi-purpose particle
detectors ATLAS \cite{Aad:2008zzm} and CMS \cite{Chatrchyan:2008aa} recorded data with an integrated luminosity of $20\,\mathrm{fb^{-1}}$.
The recorded data presents a unique opportunity to search for physics
beyond the standard model (BSM) and both experiments have interpreted their
measurements in terms of a variety of theories.\\
This work aims to briefly summarize search results in the dilepton (same and opposite flavor),
lepton$ + \met$,  dijet and ditop channel for a selected set of related BSM theories which
predict the existence of heavy gauge bosons $\Zprime$ and \Wprime, extra dimensions 
or quantum black holes.
 
\section{Theories}
\textbf{Extra dimension models} summarized in the following describe extensions 
of our spacetime with additional compactified dimension.
The related theories may lower the fundamental Planck mass $M_{D}$ to the $\mathrm{TeV}$
region, and thus solve the higgs mass hierachy problem. 
This summary focuses on the most popular theories: Randall Sundrum (RS)\cite{Randall:1999ee} and the
Arkani-Hamed, Dimopulos, Dvali (ADD) \cite{ArkaniHamed:1998rs,Antoniadis:1998ig} models. Both models provide no 
fundamental theory of quantum gravity, but are built as effective field 
theories based on classical assumptions. They use parts of the 
mathematical framework which was developed in string theory, or more precisely brane physics 
to confine SM particles to a (3+1) dimensional subspace of the ($3 + 1 +  n$) dimensional space-time\cite{ANTONIADIS1990377}.
Extra dimension theories predict a spectrum of Graviton modes (Kaluza-Klein towers) or a spectrum of
heavier copies of SM particles if they are able to propagate in the compactified additional dimensions. 

The ADD model assumes a flat spacetime.
The model parameter under study depends on the production process.
The direct production cross section depends directly on $M_{D}$,
while the virtual graviton exchange is only able to probe the UV cut-off $M_{s}$, 
which can be argued to be close to $M_{D}$.

The RS model assumes a warped space-time represented by an exponential term
in the metric $ds^2 = e^{-2kr\phi} \eta_{\mu\nu}dx^\mu dx^\nu + r_c d\phi^2$.
The cross section in these models depends on the ratio $\tilde{k}$ of the warp factor
$k$ and $M_{D}$. Several extensions of this
model exist, most notably the Bulk RS1 scenario~\cite{Agashe:2006hk}. Here the fermion and boson
fields localized near to a \TeV or a Planck brane respectively. This allows solving the flavor puzzle and 
the higgs mass hierarchy problem without introducing an additional hierarchy.
\\ \\
\textbf{Heavy Gauge Bosons }$\mathbf{\Wprime, \Zprime}$\,\,refer to heavier versions of the weak gauge bosons
and are predicted in several classes of theories.
The most studied scenario is
the sequential standard model (SSM) \cite{Altarelli:1989ff} where \Wprime \,
and $\Zprime$ bosons carry exactly the same quantum numbers and interfere 
with their SM counterparts. The $\Zprime$ is expected to decay 
flavor violating in several theories. Relevant with respect to the presented searches
are generic extensions of the SSM with additional flavor violating couplings,
which are expressed as ratios $Q_{ij}\,(i,j=e,\mu,\tau)$ of the SSM same 
flavor coupling~\cite{Murakami:2001cs} or extra dimensions where 
the mass hierachy among SM families is explained by the overlap 
of the particle wave functions if fermions and the higgs are localized 
on a higher dimensional brane~\cite{Frere:2004yu}.
Technicolor models which suggest additional gauge couplings are of special
interest in decay channels with tops where the color singlet $\Zprime$ is leptophobic and 
couples only to first and third generation quarks\cite{Harris:2011ez}. 
%~ A approach similar to the SSM is used in extended gauge models where the
%~ coupling is scaled dependent on the mass ratio $c \cdot (m_{W} / m_{\Wprime})$ 
%~ with $c$ close to unity. EGM theories predict similar results in 
%~ fermionic decay channels but produce a narrow width resonance over the 
%~ whole accessible mass range for couplings to two standard model bosons.
\\ \\
\textbf{Quantum Black Holes (QBH)} may be produced if LHC collisions take place above
a lowered fundamental Planck scale. All 
discussed models assume either the ADD or the RS model as their starting point,
but include different sets of additional assumptions. The main parameters to control the signal shape and 
cross section are the additional number of dimensions and the threshold mass $M_{th}$, which
is necessary to produce a QBH. The presented models are often referred to
by the generator which implements it. The generators used for the summarized searches are:
\begin{itemize}
	\item \texttt{CalcHEP} for flavor violating QBH decays \cite{Belyaev:2012qa}.
	\item \texttt{QBH} which uses a generic description of gravitational
bound two-body states with an non thermal QBH decay \cite{Gingrich:2009da}.
	\item \texttt{BLACKMAX} includes a wide range of black hole theories but most
relevant for the presented analyses are models comparable to the QBH generator with additional
model assumptions \cite{Dai:2007ki}.
\end{itemize}
QBH theories share an important limitation:
black hole production is expected at scales where
gravity becomes strong, and one hopes that the
extrapolation from the classical domain holds.

\section{Selected Searches with the CMS and ATLAS Experiments}
\subsection{Dilepton (same flavor)}
The dilepton channel is theoretical well understood and has been
studied by both experiments \cite{Aad:2014cka,Aad:2014wca,Khachatryan:2014fba}. Both analyses try to use a model unspecific
selection which aims to reliably select well reconstructed and isolated
pairs of electrons or muons. 
No significant deviation from the SM was observed and 
two distinct limit strategies have been used to set limits for resonant and non-resonant 
BSM signals. 

\begin{figure}[htb]
\centering
\includegraphics[height=2.1in]{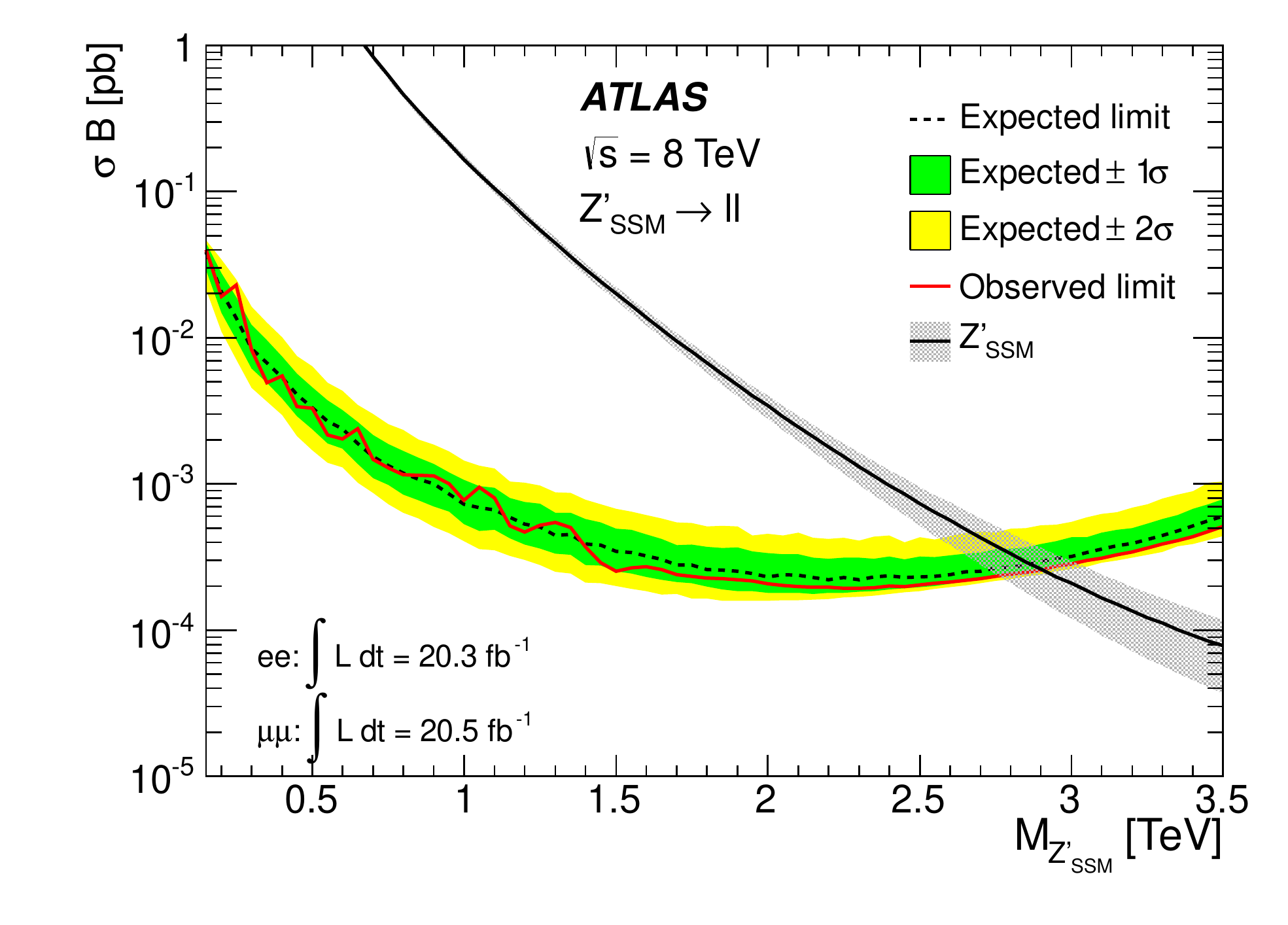}
\includegraphics[height=2.1in]{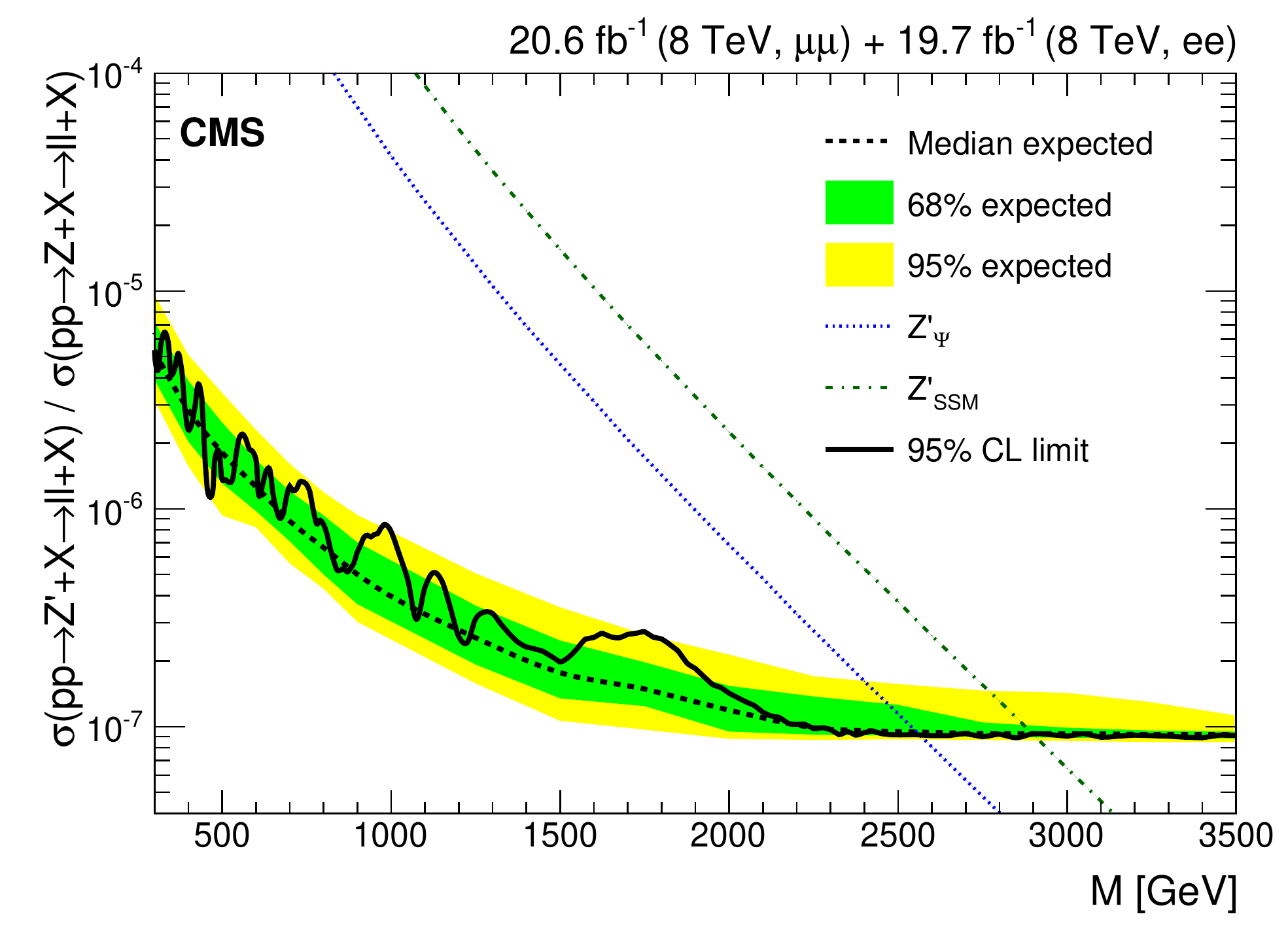}
\caption{95\% CL limits on the cross section $\times$ branching ratio $\times$efficiency dependent on the resonance mass for the
ATLAS~\cite{Aad:2014cka} (left) and CMS~\cite{Khachatryan:2014fba} (right) searches.}
\label{fig:dilepton:sflav:Zprime}
\end{figure}

The resonant searches fit smooth functions to both data and background prediction.
A set of signal shape templates with different \Zprime mass is used to construct
a background + signal hypothesis, which is compared to both data and the background only
hypothesis. 
The resulting limit on the
cross section times efficiency for a SSM \Zprime dependent on the resonance mass is shown in 
fig.~\ref{fig:dilepton:sflav:Zprime}. 
Both experiments report observed limits of 2.9\TeV on the $\Zprime_{SSM}$ mass.
The technique to derive these results differs between both experiments.
The ATLAS collaboration uses the complete spectrum with a binned likelihood approach, 
this allows gaining additional sensitivity for the studied SSM by including interference effects
outside the resonance.
The CMS collaboration has chosen a more general strategy using an unbinned likelihood approach with
a narrow width approximation. The results may be reinterpreted for any model with comparable acceptance
by simply applying a cross section ratio for the SSM $\Zprime$ and the model under investigation
within a mass window of $\pm 5 \%\, \sqrt{\hat{s}}$. This difference explains 
stronger fluctuations for CMS results in fig\,\ref{fig:dilepton:sflav:Zprime}. 

\begin{figure}[htb]
\centering
\includegraphics[height=2.1in]{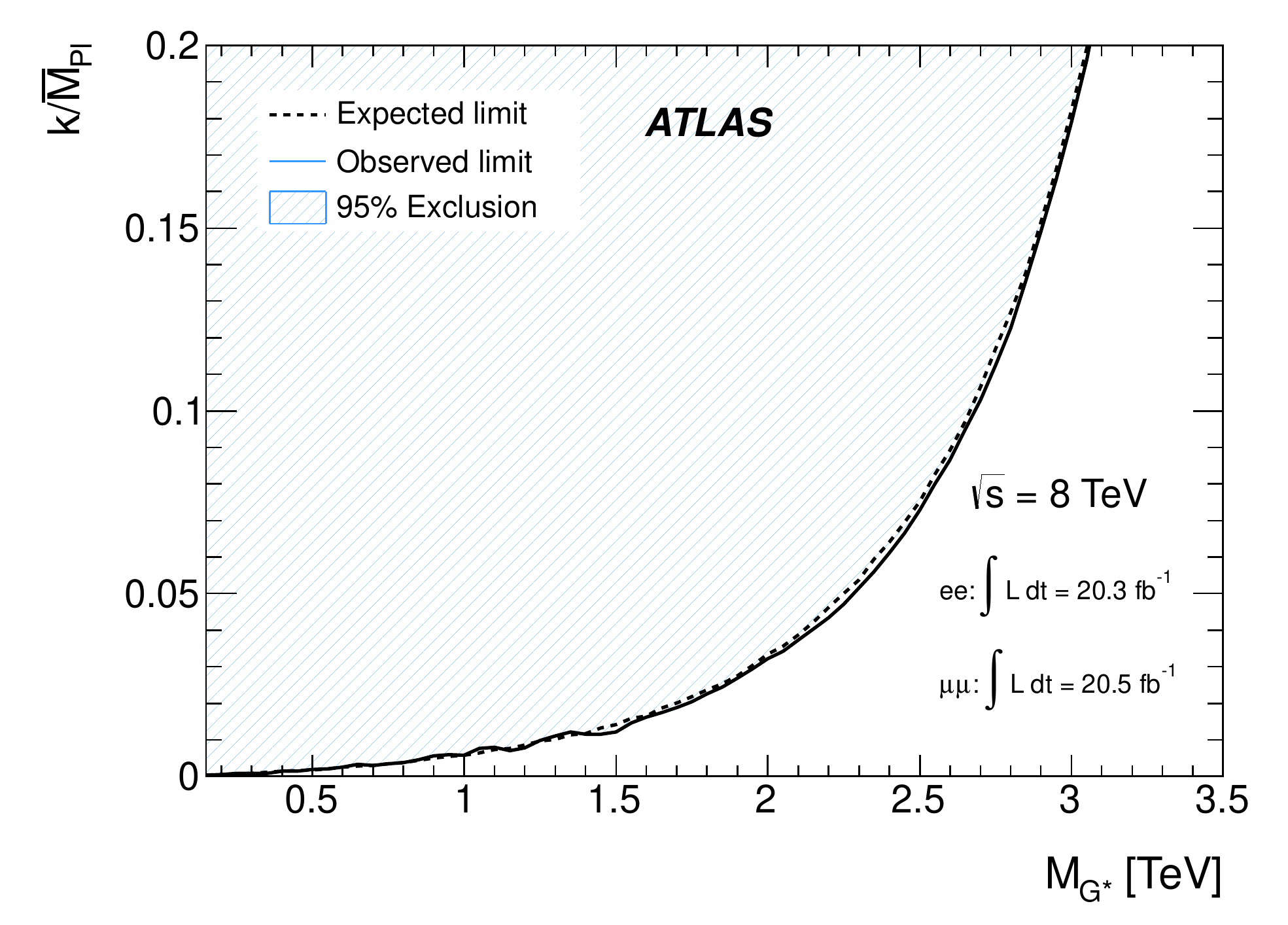}
\includegraphics[height=2.1in]{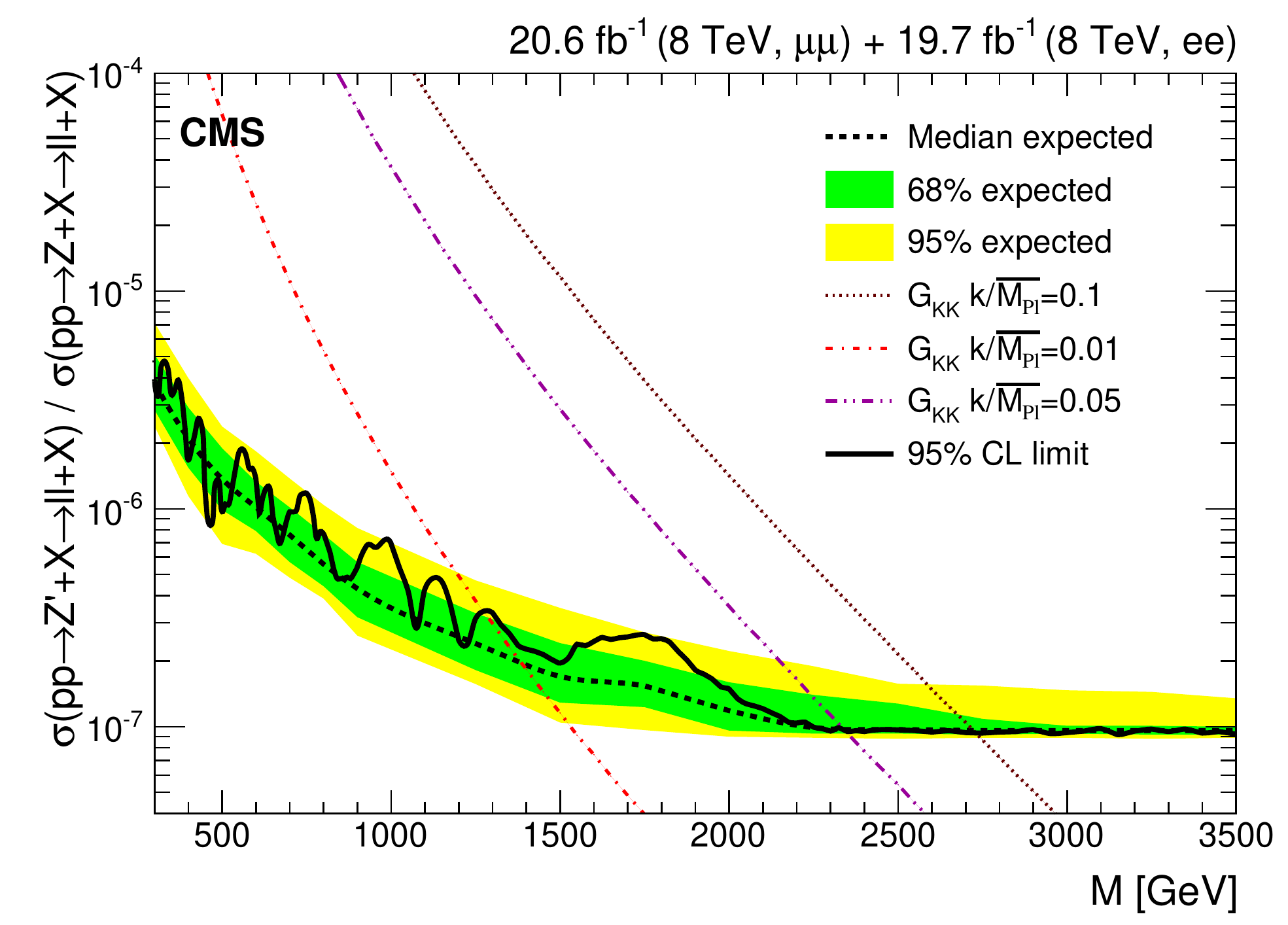}
\caption{95\% CL exclusion limits in the $k-M_{Pl}$ plane as reported by the ATLAS collaboration ~\cite{Aad:2014cka} (left)
95\% CL exclusion limits on the cross section times efficiency depending on the resonance mass for spin-2 RS Gravitons by CMS~\cite{Khachatryan:2014fba} (right).}
\label{fig:dilepton:sflav:RS}
\end{figure}

Possible signals from the lightest Kaluza-Klein Graviton mode in RS models 
serve as a benchmark model for spin 2 resonances with a modified signal acceptance.
The ATLAS results show exclusion limits in the $k-M_{Pl}$ plane, while CMS chose to
present results similar to the $\Zprime$ interpretation, see figure\,~\ref{fig:dilepton:sflav:RS}. The comparison of both CMS
limit plots show that differences in cross section between $\Zprime$ and RS Gravitons are only
visible for small resonance masses.

QBHs are expected to create an edge like resonance structure in the dilepton mass range 
close above the production threshold mass $M_{th}$. The ATLAS search uses the resonant
search strategy to derive 95\% CL exclusion limits on $M_{th}$ of  3.65\TeV for a signal
based on an ADD scenario $(n=6)$ and 2.24\TeV for a RS based scenario $(n=1)$ using the QBH generator.

\begin{figure}[htb]
\centering
\includegraphics[height=2.6in]{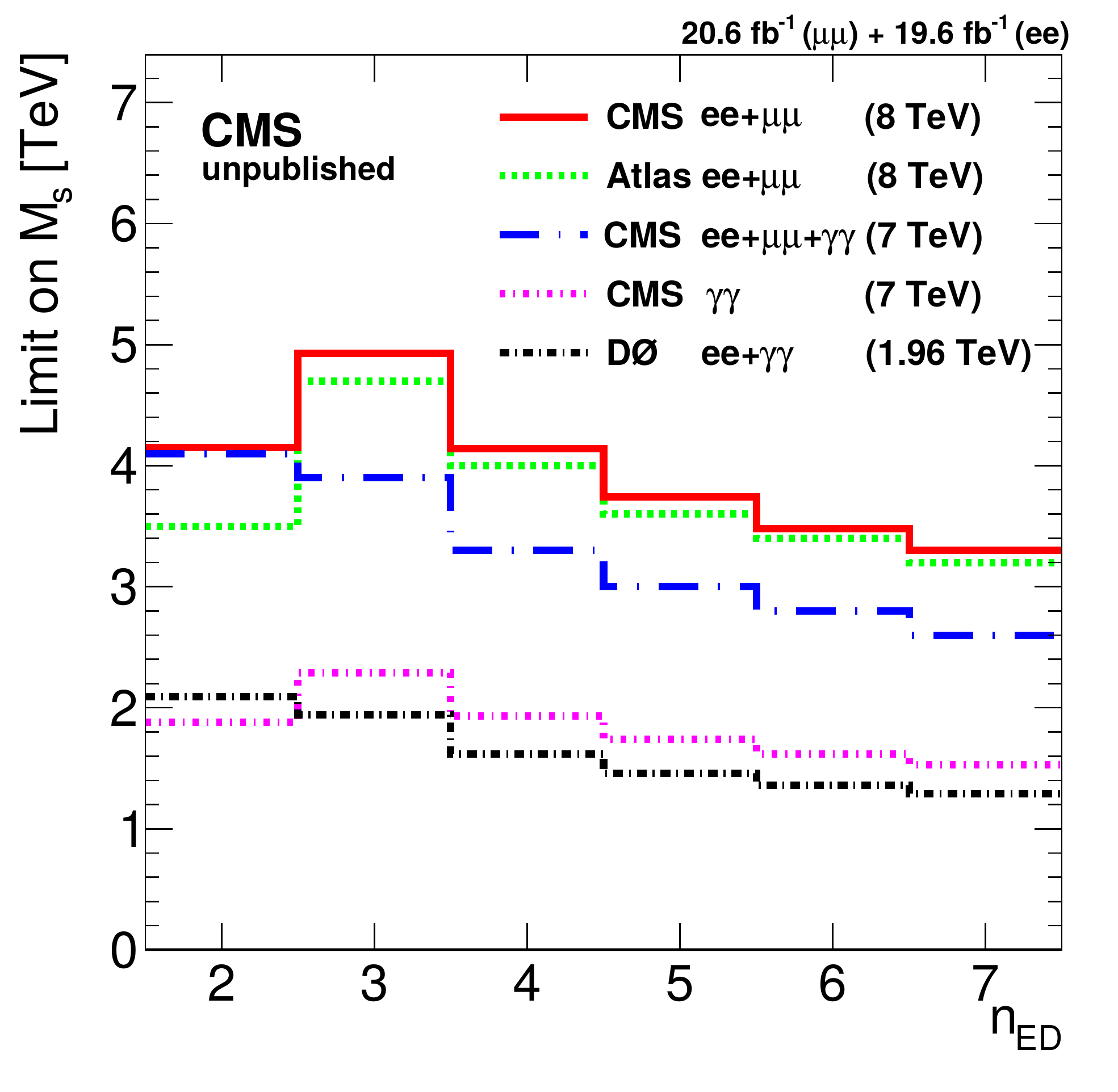}
\caption{Comparision of 95\% CL exclusion limits on the UV cutoff $M_{s}$ depending 
on the number of additional dimensions for different searches by ATLAS~\cite{Aad:2014cka}, CMS~\cite{Khachatryan:2014fba,Chatrchyan:2012kc,Chatrchyan:2011fq} and D0~\cite{Abazov:2008as}}
\label{fig:dilepton:sflav:ADD}
\end{figure}

Both experiments performed non-resonant searches using a single bin counting experiment above a
lower mass threshold, which was optimized for the best exclusion limits on 
the ADD UV cut-off $M_{s}$ at different number of extra dimensions as shown in fig.~\ref{fig:dilepton:sflav:ADD},
the observed limits reach from 4.9\TeV to 3.3\TeV for 3 to 7 additional dimensions.

\subsection{Dilepton (mixed flavor)}
Dilepton events with opposite flavor were studied by both experiments 
in the $e\mu$ channel. ATLAS has performed additional searches in $e\tau$ and $\mu\tau$
channels. Lepton flavor decays are of special interest because of the good mass resolution and
only small SM background contributions to the final states. 

Both experiments searched for $\Zprime$ bosons with additional
lepton flavor violating couplings. The ATLAS search chose the coupling $\Zprime \rightarrow e\mu,e\tau,\mu\tau$ to
be equal to SSM $\Zprime$ same flavor coupling. A binned likelihood approach was used to derive limits on the 
$\Zprime$ mass of 2.5\TeV ($e\mu$), 2.2\TeV ($e\tau$) and 2.2\TeV ($\mu\tau$) at 95\% CL. 
The CMS analysis studied an extra dimension inspired model where the coupling is
set to match existing strong bounds from $K_{L}\rightarrow e\mu$ decays. This
search concluded to be not sensitive to the $\Zprime$ model under investigation.

The quantum gravitational nature of QBHs suggest the existence of 
lepton flavor violating decays. The CMS experiment has interpreted its measurements in
terms of several QBH models implemented in CalcHEP where the threshold mass is 
set to be equal to the reduced Planck mass.
Limits at 95\% CL were set on $M_{th}$ of 2.4\TeV in a RS based scenario $(n=1)$
and 3.15\TeV to 3.63\TeV for 2 to 6 extra dimensions in an ADD based scenario.

\subsection{Lepton+$\boldmet$}
Both experiments published results for final states with one high $\mathrm{p_{T}}$
lepton and a significant amount of missing momentum in the transverse plane $\met$~\cite{ATLAS:2014wra,Khachatryan:2014tva}.
The high mass tails for this signature are dominated by off-shell SM $W$ production.
Single lepton triggers with transverse momentum thresholds for electrons (muons) of 
$\mathrm{p_{T}}>120\GeV \, (\mathrm{p_{T}}>40\GeV)$ and $\mathrm{p_{T}}>80\GeV \, (\mathrm{p_{T}}>40\GeV)$ have been used by ATLAS
and CMS respectively.
Events with additional well reconstructed same flavor leptons with $\mathrm{p_{T}}>20\GeV$ are 
discarded in the ATLAS analysis while CMS uses $\mathrm{p_{T}}>35\GeV$ for electrons and $\mathrm{p_{T}}>25\GeV$ for muons.
The transverse mass $M_{T}=\sqrt{2\mathrm{p_{T}}^{\mathit{l}}\met\left(1 - \cos[\Delta \phi(\vec{\mathrm{p_{T}}}^{\mathit{l}},\vec{\met})]\right)}$ is used
as the main observable for $\Wprime$ searches.

Additional final state specific kinematic cuts distinguish both searches: ATLAS adjusts the lower
threshold for $\met$ to the trigger $\mathrm{p_{T}}$ thresholds for each flavor; CMS applies a back-to-back
cut $\Delta\phi(l,\met) > 2.5$  and requirements on the $\mathrm{p_T}$-\met ratio: $0.4 <\mathrm{p_{T}} / \met < 1.5$, both
cuts should reflect that BSM paricles are produced in a balanced recoil at leading order.

\begin{figure}[htb]
\centering
\includegraphics[height=2.3in]{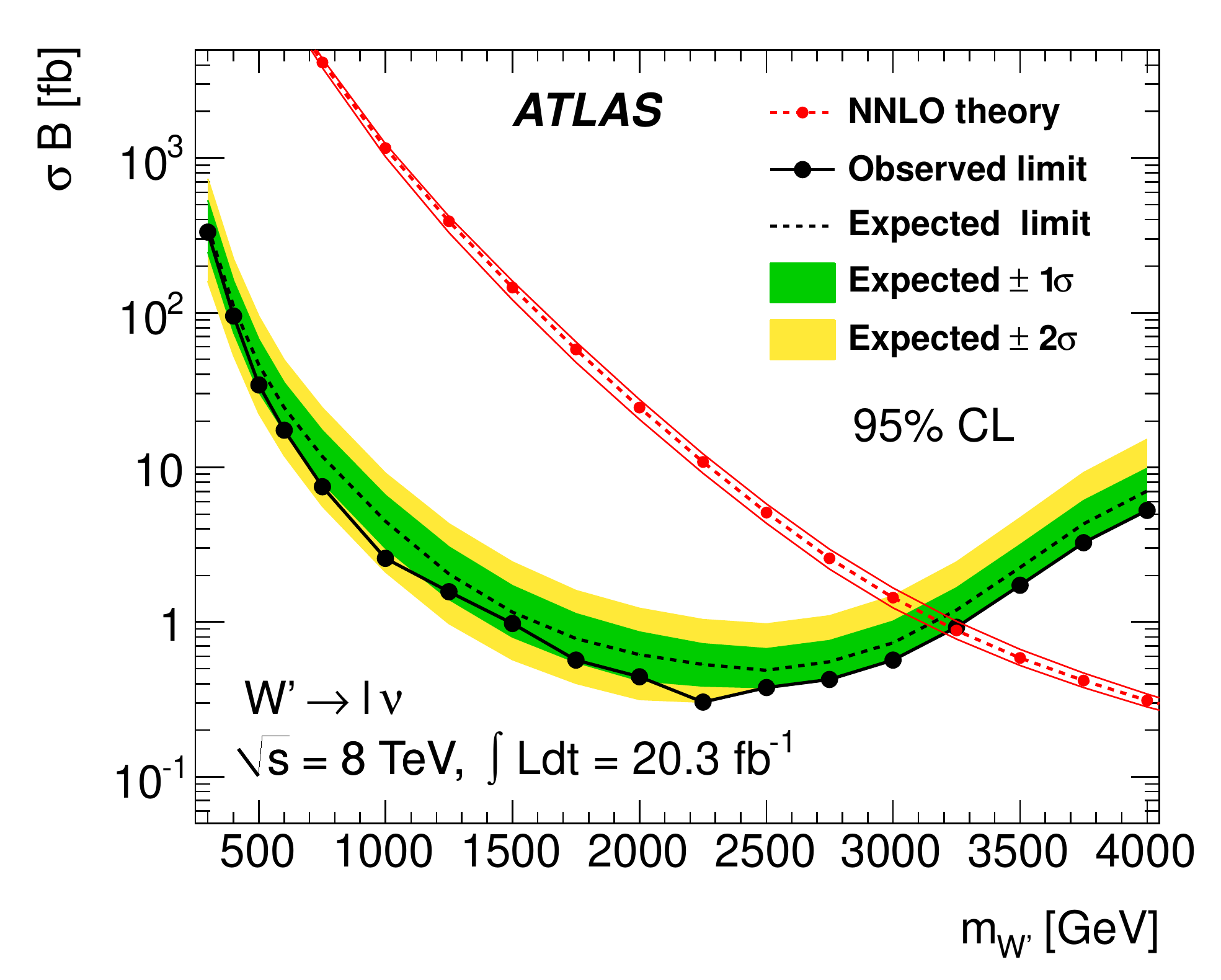}
\includegraphics[height=2.3in]{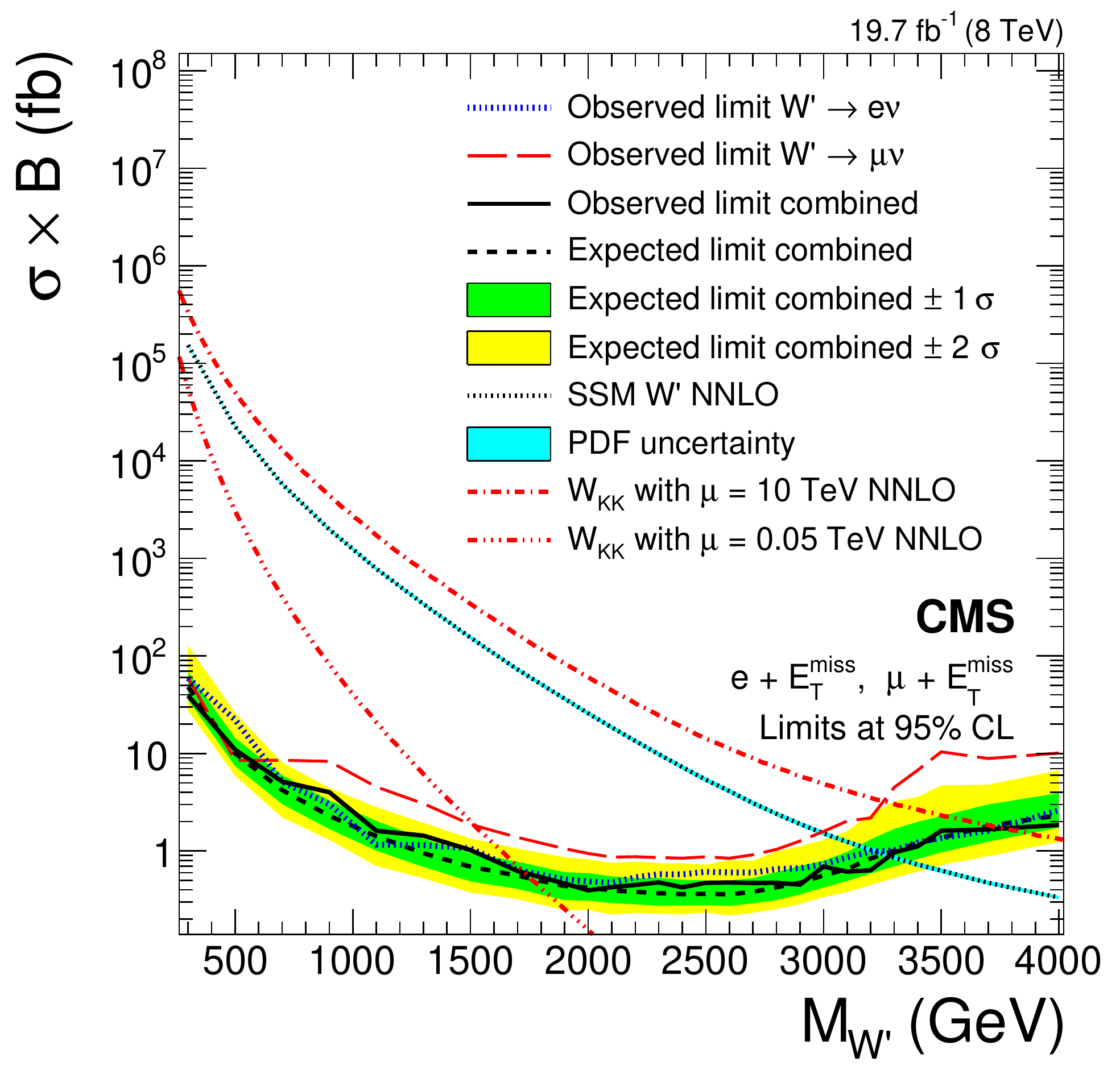}
\caption{95\% CL exclusion limits on branching ratio times cross section depending on the \Wprime mass for
ATLAS~\cite{ATLAS:2014wra}\,(left) and CMS~\cite{Khachatryan:2014tva}\,(right)}
\label{fig:monolepton:wprime}
\end{figure}

ATLAS and CMS report lower limits of 3.2\TeV and 3.3\TeV on the $\Wprime$\, mass at 95\,\%CL.
Different statistical procedures were used to derive the limits; 
ATLAS uses a single bin counting experiment above a varying lower threshold on $M_{T}$.
Cross section limits are calculated based on an optimized threshold for each
considered \Wprime\, mass, see fig.~\ref{fig:monolepton:wprime}. The CMS analysis used
an shape based template fit similar to the resonant ATLAS search in the dilepton channel, see fig.~\ref{fig:monolepton:wprime}.
CMS has also reported limits based on single bin counting experiments above varying mass thresholds 
but did not use this approach for the \Wprime\, interpretation.

\subsection{Dijet}
Final states with two high-$\mathrm{p_{T}}$ jets profit from a large cross section at hadron colliders
like the LHC and enough events were collected to extract shape information up to several \TeV.
Both experiments add additional requirements on the dijet event kinematics in their search~\cite{Aad:2014aqa,Khachatryan:2015sja}.
A separation in (pseudo)-rapidity
between the jets with the highest $\mathrm{p_{T}}$ of $\Delta y <  0.6$ and 
$\Delta \eta <  0.65$ is used by ATLAS and CMS respectively.
ATLAS used so called pre-scaled triggers where only a fixed fraction of
all events is saved. This allows collecting data with lowered trigger requirements
and decreases the lower limit for searches in the dijet mass distribution to
$m_{jj}>250\GeV$ compared to the CMS analysis with $m_{jj}>890\GeV$.
Both experiments use smooth fit function to estimate the background expectation 
from data and compare it to signal templates using a binned likelihood approach.
Lower limits on particle masses for SSM \Zprime,\,\Wprime\, and Kaluza-Klein Gravitons
in the RS model with $n=1$ are listed in table~\ref{tab:dijet:primes}.
\\ \\
Dijet events also represent the most sensitive channel for QBH searches, and
many QBH models predict that the produced BH decays primarily to dijet pairs~\cite{Calmet:2008dg}.
Lower limits on $M_{th}$ were set by both experiments using the model implemented in the \texttt{QBH} generator.
ATLAS has set $M_{D}=M_{th}$ and reported 5.7\TeV while CMS kept both variables as free parameters and
found a limit of 5.8\TeV for $M_{pl}=5\TeV$. Additional bounds on a related model implemented in \texttt{BLACKMAX} 
were set by the ATLAS experiment of $M_{th}<5.6\TeV$ where $M_{th}$ is again set to be equal to the reduced Planck mass $M_{D}$.

\begin{table}[t]
\begin{center}
\begin{tabular}{l|ccc}  
[\TeV]	&  \Wprime &  \Zprime & $G_{KK}$(RS)\\ \hline
 ATLAS   &   2.5     &        &    \\
 CMS  	&  2.2     &     1.7     &  1.6 \\ \hline
\end{tabular}
\caption{95\%\,CL lower mass limits on the SSM \Wprime, $\Zprime$ and $G_{KK}$ (RS $n=1$) as reported by ATLAS~\cite{Aad:2014aqa} and CMS~\cite{Khachatryan:2015sja} for the dijet channel.}
\label{tab:dijet:primes}
\end{center}
\end{table}

\subsection{Ditop}
\begin{figure}[htb]
\centering
\includegraphics[height=1.5in]{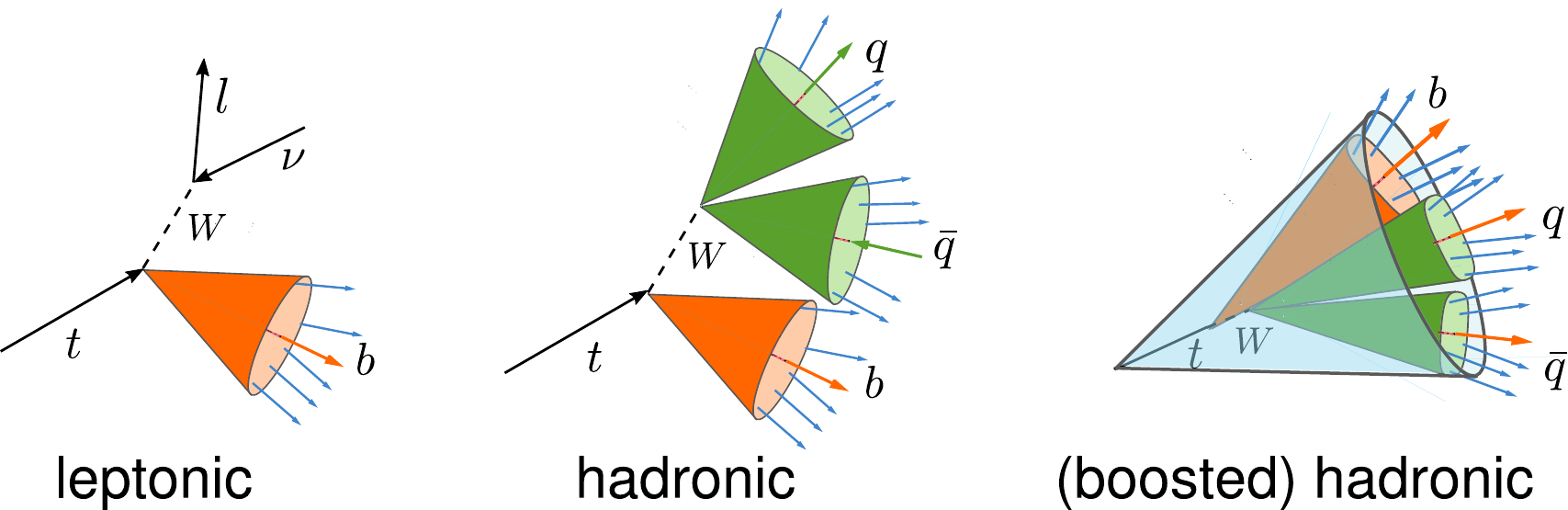}
\caption{ Graphical representation of the possible decay modes for a single top quark.}
\label{fig:ditop:decaymodes}
\end{figure}

The analysis of ditop final states by ATLAS and CMS~\cite{Aad:2015fna,CMS:2015nza} has significantly
increased its sensitivity by employing new analysis strategies for the
reconstruction of boosted top decays, and the subsequent top identification
via so called top-tagging techniques. Each of the two tops decays
either leptonically or hadronically, the hadronic decays can be further split into
a resolved and boosted topology, see fig.~\ref{fig:ditop:decaymodes}.
The combination of these decay modes for both tops results in the
ditop decay modes: leptonic-leptonic, leptonic-hadronic, leptonic-hadronic(boosted), hadronic-hadronic
hadronic(boosted)-hadronic(boosted).
ATLAS restricted its analysis to the most sensitive combination with one leptonic and one
hadronic decay for the models under investigation, CMS analyzed all possible decay modes and
combined the measurements for the final result.

\begin{figure}[htb]
\centering
\includegraphics[height=2.1in]{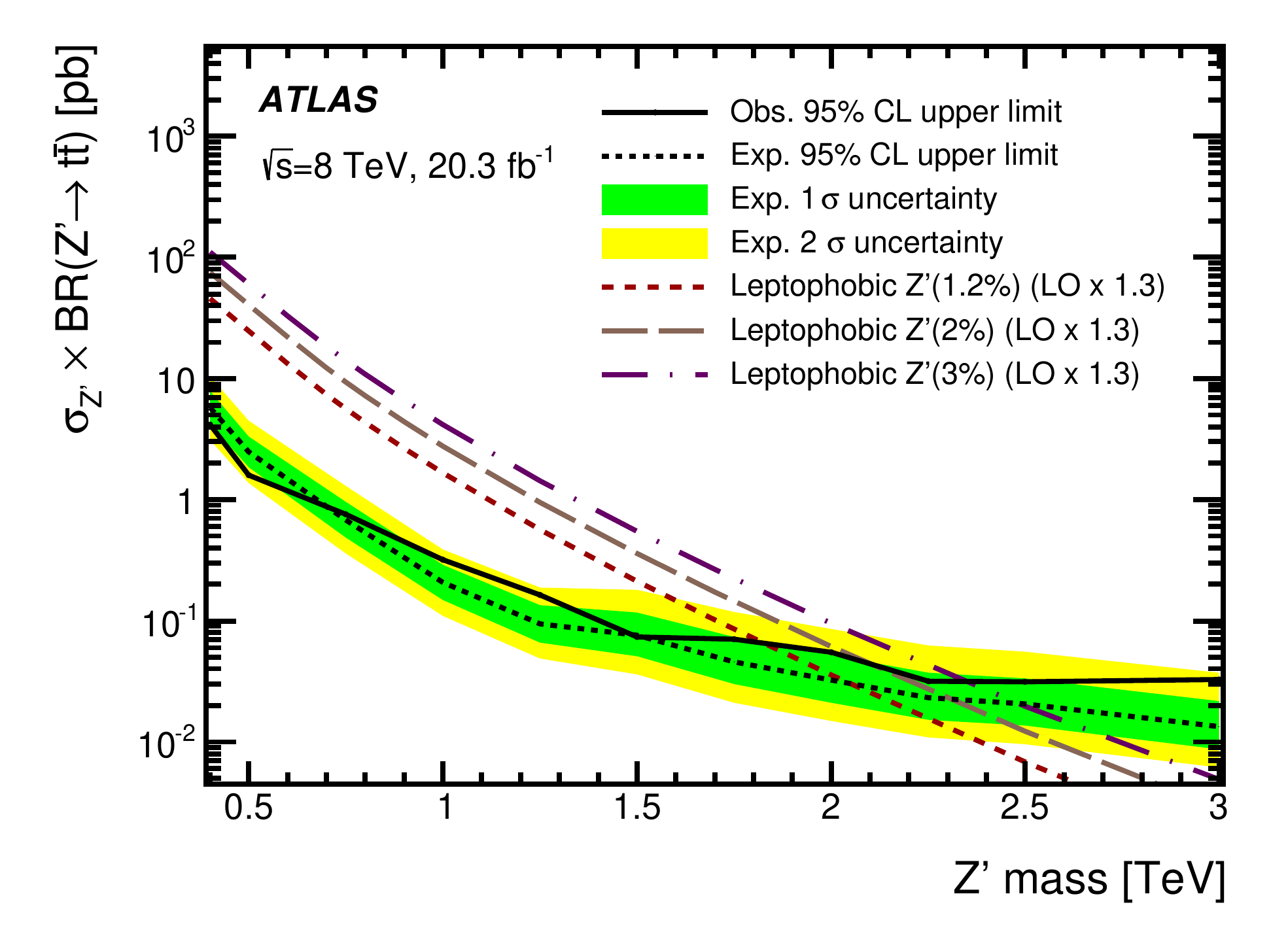}
\includegraphics[height=2.2in]{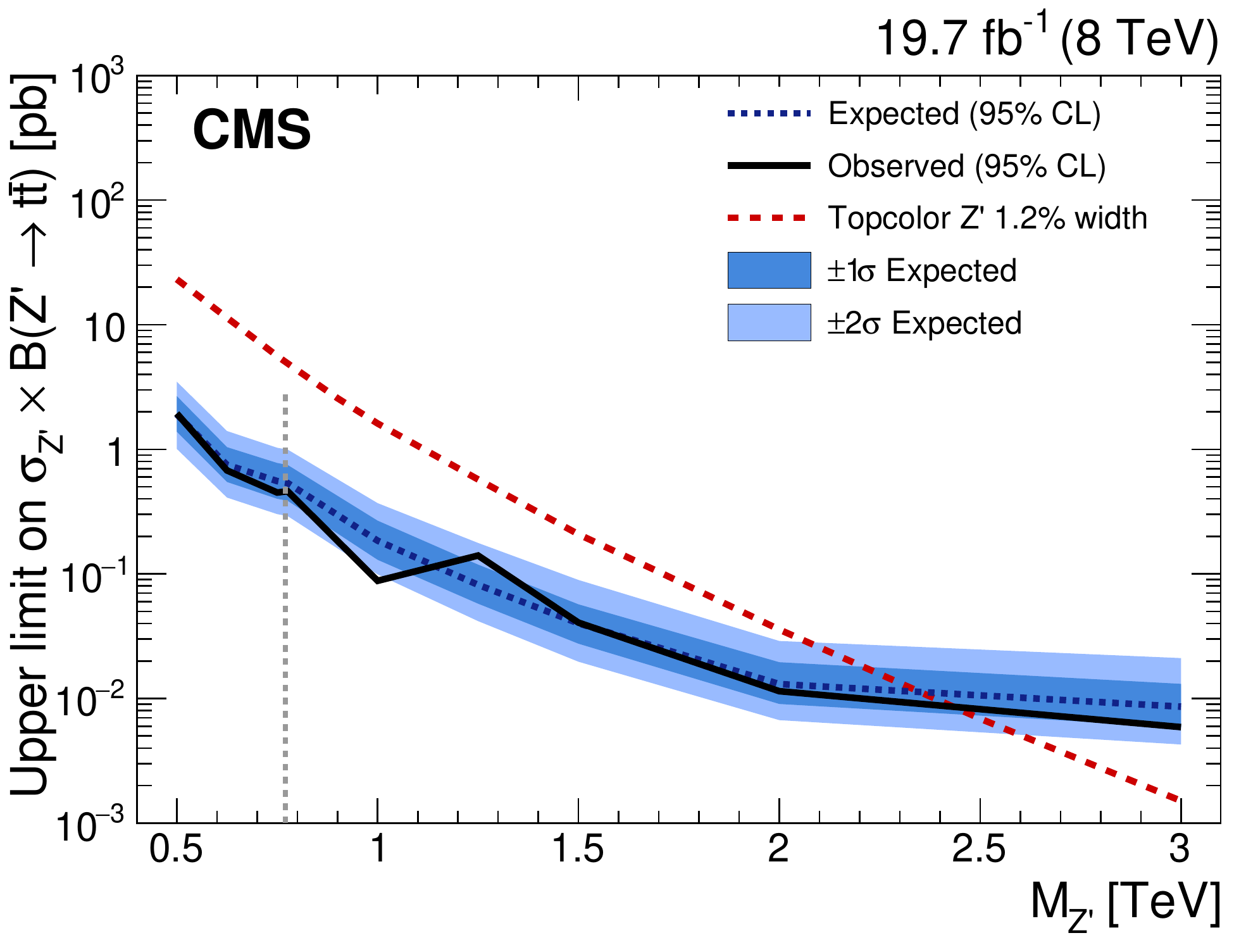}
\caption{95\% CL exclusion limits on the branching ratio$\times$cross section dependent on the \Zprime mass for
ATLAS~\cite{Aad:2015fna}\,(left) and CMS~\cite{CMS:2015nza}\,(right)}
\label{fig:ditop:zprime}
\end{figure}

Limits have been set on the $\Zprime$ mass based on topcolor models as described in~\cite{Harris:2011ez} where the 
coupling to lighter quarks is suppressed:
ATLAS  and CMS found lower limits of 1.8\TeV and 2.4\TeV with a width of 
1.2\% and 1\% of the $\Zprime$ mass respectively, see figure\,~\ref{fig:ditop:zprime}. \\
The Bulk RS1 model expects a suppression in production and decay for lighter quarks.
This leaves $t\bar{t}$ final states as the most promising channel
to probe the production of Kaluza-Klein gluons $g_{KK}$ at the LHC~\cite{Agashe:2006hk}.
ATLAS and CMS report lower limits on the mass of the lightest Kaluza-Klein mode of the gluon
$g_{KK}$ of 2.2\TeV and 2.8\TeV respectively.

\section{Conclusion}
ATLAS and CMS have both performed a large number of searches for the presented theories and it should be 
emphasized that this summary reports only on a small subset of all searches. 
No significant evidence for physics beyond the standard model has been reported.
A comprehensive list of all searches for new physics related to this talk are
constantly updated online ( 
\textbf{ATLAS:}  \href{https://twiki.cern.ch/twiki/bin/view/AtlasPublic/ExoticsPublicResults}{ExoticsPublicResults}
\textbf{CMS:}   \href{https://twiki.cern.ch/twiki/bin/view/CMSPublic/PhysicsResultsEXO}{PhysicsResultsEXO},
                \href{https://twiki.cern.ch/twiki/bin/view/CMSPublic/PhysicsResultsB2G}{PhysicsResultsB2G} ).
The reach for most of the presented analysis is limited by the probable phase space.
The recent restart of the LHC at a center of mass energy of $\sqrt{s}=13\TeV$ will increase the discovery reach
for most theories with a fraction of the recorded integrated luminosity at $8\TeV$.  

%~ \\ \\
%~ \textbf{Boosted single top selection}
%~ ATLAS uses large anti-kt jets (R=1.0) and searches for smaller inclusive $k_{t}$ subjets (R=0.3) in it.
%~ This approach works due to the good geometrical resolution in ATLASs calorimeters.
%~ Subjets wich do not carry a significant amount of the large jet pt are removed to subtract effects from pileup and
%~ at least one b jet is expected to be close to the large jet or lepton.

%%%%%%%%%%%%%%%%%%%%%%%%%%%%%%%%%%%%%%%%%%%%%%%%%%%%%%%%%%%%%%%%%%%%%%%%%
%%
%%   use this format to include an .eps figure into your paper
%%
%~ \begin{figure}[htb]
%~ \centering
%~ \includegraphics[height=1.5in]{magnet}
%~ \caption{Plan of the magnet used in the mesmeric studies.}
%~ \label{fig:magnet}
%~ \end{figure}
%~ %%%%%%%%%%%%%%%%%%%%%%%%%%%%%%%%%%%%%%%%%%%%%%%%%%%%%%%%%%%%%%%%%%%%%%%%%%%
%~ 

%%%%%%%%%%%%%%%%%%%%%%%%%%%%%%%%%%%%%%%%%%%%%%%%%%%%%%%%%%%%%%%%%%%%%%%%%
%%
%%   use this format to include a LaTeX table  into your paper
%%
%~ \begin{table}[t]
%~ \begin{center}
%~ \begin{tabular}{l|ccc}  
%~ Patient &  Initial level($\mu$g/cc) &  w. Magnet &  
%~ w. Magnet and Sound \\ \hline
 %~ Guglielmo B.  &   0.12     &     0.10      &     0.001  \\
 %~ Ferrando di N. &  0.15     &     0.11      &  $< 0.0005$ \\ \hline
%~ \end{tabular}
%~ \caption{Blood cyanide levels for the two patients.}
%~ \label{tab:blood}
%~ \end{center}
%~ \end{table}
%%%%%%%%%%%%%%%%%%%%%%%%%%%%%%%%%%%%%%%%%%%%%%%%%%%%%%%%%%%%%%%%%%%%%%%%%%%

\Acknowledgements
I am grateful to Serguei Petrouchanko, Johannes Haller, Tobias Golling and Koji Terashi	for their
helpful input during the preparation of my conference contribution.
I thank CERN and the ATLAS and CMS collaborations for their great work operating the LHC and for providing the
results for this summary.

\end{document}